\documentclass{iaus}
\usepackage{graphicx}

\title[JD 11.~~Grain Chemistry of Interstellar Deuterium] 
{The Effects of Subsurface Chemistry in the Grain Mantles
on the Deuterium Chemistry in Molecular Clouds}

\author[Juris Kalv{\-a}ns \& Ivar Shmeld]   
{Juris Kalv{\-a}ns$^{1,2,3}$
 \and Ivar Shmeld$^{1,2,4}$}

\affiliation
{$^1$Ventspils International Radioastronomy Centre of Ventspils University College \\ In$\mathrm{\check{z}}$enieru iela 101, Ventspils, Latvia \\[\affilskip]
$^2$Institute of Astronomy, University of Latvia \\ Raina 19, Riga, Latvia \\[\affilskip]
$^3$email: {\tt kalvans@lu.lv}
$^4$email: {\tt ivarss@venta.lv}}

\pubyear{2011}
\volume{280}  
\pagerange{119--126}
\jname{The Molecular Universe}
\editors{A.C. Editor, B.D. Editor \& C.E. Editor, eds.}
\begin{document}

\maketitle

\begin{abstract}
The deuterium enrichment in molecules in dark molecular cloud cores and star-forming regions is usually attributed to gas-phase chemistry. Here we examine the effects of surface and mantle chemical reactions on the deuteration of species. We use a simple kinetic chemistry model that includes gas, surface and mantle pore phase reactions of deuterated species. The mantle is assumed to be partially reactive due to pores with sufficient surface area for chemical reactions, that are continuously transformed by cosmic-rays. Calculation results show that surface reactions generally enhance the deuteration for at least several molecules. However, once they are buried and become mantle molecules, they lose their deuteration over a timescale of 10 million years due to processes in the mantle. If deuterated species in young star-forming regions come from grain mantles, a cautious conclusion is that the freeze-out of molecules, perhaps, should not occur more than 10 Myr before the mantle evaporates to the gas phase.

\end{abstract}

\firstsection 
\section{Introduction}

Current models of molecule formation in dense, cold molecular cloud cores include two chemically active phases -- the gas and the dust grain surface. In the paper by Juris Kalv$\mathrm{\bar{a}}$ns \& Shmeld (\cite{01}), further referenced as Paper1, the existence of a third active phase was proposed, i.e. the surface of cavities inside the grain mantle. It is a development of the three-phase model originally proposed by Hasegawa \& Herbst (\cite{02}).

There were several key elements in astrochemical modelling proposed in Paper1. First, it is assumed that cosmic rays passing through the grain periodically rearrange the structure of the loosely-bound mantle, creating new cavities and destroying others. Thus, a large part of the mantle molecules are cycled and transformed through the active cavity-surface phase. Second, the molecules on the grains (mantle and surface) are subjected to photodissociation from cosmic-ray induced photons, much like in the gas phase. Third, hydrogen diffusion takes place from the surface to the cavities and in the opposite direction.

The deuterium accumulation in molecules containing heavy elements is a well-established observation fact (see, e.g. Solomon \& Woolf \cite{12}, Langer et al. \cite{13}). These molecules evaporate into the gas phase when the core eventually shrinks further and heats, when hot core forms. We examine the effect of the inclusion of chemically active mantle and surface phases in a simple astrochemistry model on the deuterium enrichment in heavy molecules.

\section{The physical model}
\label{phys}

A steady-state local thermodynamic equilibrium model for a cloud in thermodynamic equilibrium is used. The integration time for kinetic reactions in the model was taken to be $10^{13}$ to $10^{16}s$. Like in Paper1, it is assumed that a dark, dense molecular cloud core with density $n_{H}=10^{5}cm^{-3}$ forms in approximately a million years since the beginning of the collapse of the cloud. The icy mantle of a grain accretes in a similar timescale. The gas temperature in the core is taken to be 15\emph{K}, the dust grain temperature -- 10\emph{K}. Surface and mantle molecules are treated as solid species uniformly dispersed in volume with abundances expressed in $cm^{-3}$. All rate coefficients for physical phase-transformations in Sect. ~\ref{phys} have the dimension of $s^{-1}$.

\subsection{Surface}
\label{surface}

The surface is defined as the first few (two in this paper) layers of molecules on top of the mantle that are exposed to surface reactions and desorption processes. The rate coefficients for accretion of molecules onto grains (Willacy \& Williams, \cite{03}), thermal evaporation and desorption by grain heating by cosmic rays (Hasegawa \& Herbst, \cite{05}), desorption by cosmic-ray induced photons (Prasad \& Tarafdar, \cite{06}, Willacy, Williams, \cite{03}), desorption of surface molecules by energy released by $\mathrm{H_{2}}$ molecule formation on grains (Roberts et al., \cite{07}) all are calculated according to Paper1. The species desorbed for the latter mechanism are those with binding energies $E_{b} \leq 1210K$ (binding energies from Aikawa et al. \cite{08}).

\subsection{Mantle}
\label{mantle}

The mantle forms slowly from buried surface molecules. It is assumed that the mantle molecules are not permanetly locked in their place. Instead, the mantle is turned completely inside-out by cosmic-rays in a time related to the cosmic-ray intensity. The mantle molecules are returned to surface with a rate coefficient
   \begin{equation}
   \label{dig-up}
k_{dig-up}=t^{-1}_{cr}\times v^{-1},
   \end{equation}
where $t_{cr}=3.16\times10^{13}$ (Hasegawa \& Herbst \cite{05}) is the average time between two successive strikes by Fe cosmic-ray nuclei and \textit{v}=8 is the assumed number of strikes needed to return all mantle molecules to the outer surface for some time. Thus we deliberately ignore the division of the mantle in molecule layers of various depths, since the mixing of the mantle is an essential feature in our model. The mantle formation coefficient $k_{mantle}=15k_{dig-up}$ is then adjusted such that it produces approximately a 15:1 ratio of mantle and surface molecule abundances which equals 2 monolayers of ``surface'' molecules in a 30-monolayer thick mantle. These rate coefficients ensure that more than 90\% of heavy species are the solid phase after time of order 1 million years.

The mantle itself consists of two phases - cavity surface phase, where different reactions take place, and inactive volume phase. The cavity molecules and an additional, 2\% more of mantle volume molecules (next to a cavity) are subjected to photodissociation from cosmic-ray induced photons that produces reactive species on cavity surfaces. We assume that the products from the dissociation of these additional molecules enter the cavity or become exposed to the cavity surface and thus are reactive. The photo-dissocioation of molecules in the remainder of the mantle lead to inactive products. The yields of photodissociation of the various phases are summarised in Table~\ref{tab-phases}.

The molecules in the mantle are cycled through the active cavity phase with a rate coefficient
   \begin{equation}
   \label{act}
k_{act}=t^{-1}_{cr}\times u^{-1},
   \end{equation}
where $u$ is the number of strikes needed to fully reorganize the grain mantle, i.e. to expose to an inner surface an amount of molecules equivalent to the total number of molecules per mantle. In this paper we rather arbitrary assume that $u=0.5v=50$. Intact molecule activation by direct cosmic-ray hits and dissociation by CR-induced radiation of molecules nearby to the cavities are the two processes that change the placement and the shape of the cavities in the mantle. The molecules are returned to the passive mantle volume phase according to rate coefficient
   \begin{equation}
   \label{inact}
k_{inact}=t^{-1}_{cr},
   \end{equation}
i.e., all cavities are resurfaced upon the next Fe cosmic-ray impact.

\begin{figure}[b]
 \vspace*{-1.0 cm}
 \begin{center}
 \includegraphics[width=5.5in]{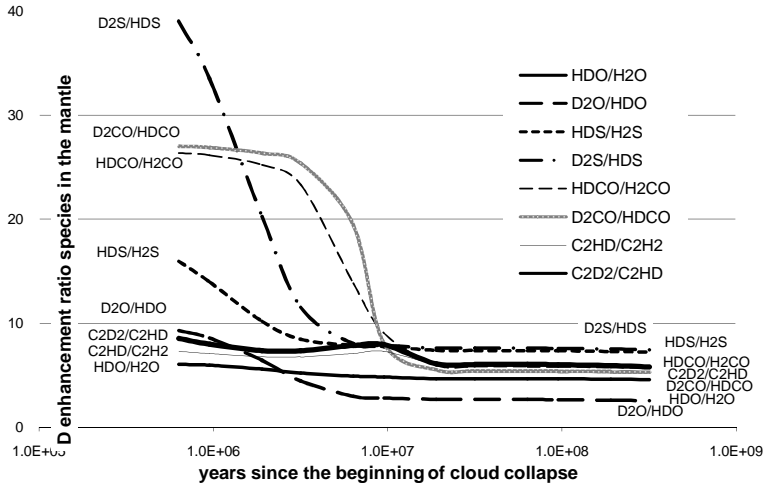} 
 \vspace*{-11.0 cm}
  \caption{The modelled evolution of the deuterium to hydrogen ratio in selected heavy species in the mantle on grains, divided by the amount of total hydrogen atoms in the molecule. The equilibrium between gas and solid phases (accretion and desorption) has already been reached. The time is in years after the collapse of the cloud and the formation of the dark, dense molecular core.}
   \label{graf}
\end{center}
\end{figure}

For hydrogen diffusion each of the three solid phases play an important role. The key element here is that according to the conclusions by Strauss et al. (\cite{10}), hydrogen almost always resides on a surface, and H and $\mathrm{H_2}$ residing in the lattice is only a rare intermediate state. First, hydrogen atoms and molecules accrete on the grain (mantle) surface. They also reside on inner surfaces in the cavity phase and in the more densely packed amorphous volume phase - in microcavities. A microcavity is too small to provide enough surface and reactants for chemical reactions, but it is large enough to provide a stable surface for one or few hydrogen atoms/molecules to reside on. Thus the two mantle phases also represent two kinds of cavities, when regarding the diffusion of hydrogen in our model.

The rate coefficient of hydrogen diffusion through the mantle is calculated according to
   \begin{equation}
   \label{hdiff1}
k_{diff}=D P_{diff} L,
   \end{equation}
where \textit{D} is the diffusion coefficient, $P_{diff}$ is the probability for diffusing species to find an appropriate surface, further referenced as \textit{P} for short. $L$ is the length of the diffusion path (we take it always $1\times10^{-6} cm$), approximately half the thickness of the mantle. $D_{\mathrm{H}}=2.5\times10^{-21} cm^{2} s^{-1}$ from Awad et al. (\cite{11}) and $D_{H_2}=5.9\times10^{-8} cm^{2}s^{-1}$, as estimated from the data by Strauss et al. (\cite{10}) in Paper1. No literature data could be found on the diffusion coefficients of D, HD and $\mathrm{D_2}$ at cryogenic temperatures in an environment similar to interstellar ices. However, the exact value of this parameter does not affect the modelling results for heavy molecules. We assumed the values $10^{-7}, 10^{-6}$ and $10^{-8}$ of diffusion coefficients for D, HD and $\mathrm{D_2}$ respectively, relative to the diffusion coefficients of their respective protium analogues. These are, perhaps, the lowest feasible values, when it is assumed that the diffusion coefficient depends on the activation energy barrier for diffusion (Lipshtat et al. \cite{28}).

The hydrogen diffusion probabilities to the outer surface $P_S$, to cavities $P_C$, and to the volume of the mantle (microcavities) $P_V$ are dependent on the relative surface area (i.e., the proportion of molecules) and the accesibility of these phases and are summarised in Table~\ref{tab-phases}.

\begin{table}
\caption{Properties of the three solid phases (outer surface, cavity surface and volume that includes microcavities): the desired proportion of the mantle consisting if the specific phase, hydrogen diffusion probability to it, and the total yield of dissociation by CR-induced photons.}
\label{tab-phases}
	\centering
\vspace*{0.5 cm}
		\begin{tabular}{cccc}
\hline\hline
Phase&\% of the mantle&$P_{diff,H}$&$Y_{dis,tot}$\\
\hline
Surface&7&0.5&0.1\\
Cavity&2&0.04&0.1\\
Volume&91&0.2&0.05\\
\hline
		\end{tabular}
\end{table}

\section{The chemical model}
\label{chem}

\begin{figure}[b]
 \vspace*{-1.5 cm}
\begin{center}
 \includegraphics[width=6.0in]{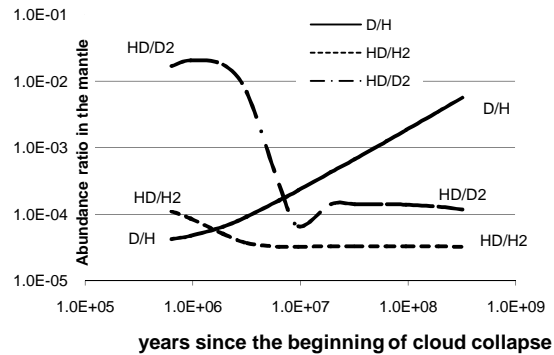} 
 \vspace*{-14.5 cm}
 \caption{The modelled evolution of atomic and molecular deuterium to hydrogen ratio in the icy mantle of dust grains.}
   \label{grafH}
\end{center}
\end{figure}

The UMIST udfa06 dipole astrochemistry database (Woodall et al. \cite{22}) is used to provide the gas-phase chemical reaction set. The elements permitted are H, D, He, C, N, O, Na, Mg, S, Fe. We adopted elemental abundances provided by Jenkins (\cite{23}) except for deuterium with relative abundance $2\times10^{-5}$ taken from Prodanovic et al. (\cite{24}).
Deuterium is involved in the reaction network keeping the approach by Rodgers \& Millar (\cite{25}). That is, the rate of a gas-phase reaction is equal for deuterated and non-deuterated species and statistical branching ratios for reaction outcomes are assumed. For grain and cavity surface reactions, the reaction rate of deuterium species is modified by the mobility of molecules, calculated according to Hasegawa et al. (\cite{26}). In total, the model includes 491 chemical species with more than 14000 reactions. No species with more than 4 heavy atoms and 4 hydrogen atoms are included, with the sole exception of $\mathrm{CH_{5}^+}$ and its hydrogen isotopologues. Of these, 186 neutral species are refractory, accreting and participating in the chemistry on the grains. An approximation is made that in a molecule there is no distinction between H or D atoms attached to different heavy atoms. For example, methanol is treated as $\mathrm{CH_4O}$, not $\mathrm{CH_3OH}$. This approximation eases calculations at the cost of only little loss of information because the actual fate of different hydrogen atoms in a reaction usually is unknown anyway. In addition, we include the deuterium exchange reactions from Roberts \& Millar (\cite{19}). The reaction set is available from the authors.

Solid phase binary reactions (Hasegawa \& Herbst, \cite{05}) and the dissociation by CR-induced photons of the mantle species (Woodall et al., \cite{12}) with dissociation yields specified in Table~\ref{tab-phases} are treated according to Paper1.

\section{Results and Discussion}
\label{results}

We focus on investigating the enrichment of deuterium in molecules in the mantle, where most of the heavy species reside in dark cores. Generally, most of the solid species included in the model show weak deuteration, that further decreases with time. Several species with the highest D/H ratio, among others, are shown in Fig.~\ref{graf}.

Several remarks worth noting emerge from the examination of calculation results. First, the abundance of deuterated species (relative to molecule deuteration in gas phase) is increased by grain surface reactions. These molecules are then included in the mantle. The continuous exchange of hydrogen between the surface and the mantle in combination with the more passive chemistry in the mantle then reduces the deuteration effects. The D/H ratio in mantle molecules is driven to similar to the surrounding gas in time of order $10^7$ years (Fig.~\ref{graf}). The diffusion coefficient for deuterium (most importantly for the HD form due to its fastest diffusion rate) is still too high to enable an accumulation of deuterium in the mantle. Instead, an equilibrium is reached that generally erases the deuterium enhancement in molecules reached by gas and surface phase chemistry. Thus, a cautios conclusion might be that efficient deuteration processes will have no effect if the freeze-out of molecules occurs more than $10^7$ years before the formation of a hot core for species released in gas phase, because they will be erased by mantle chemistry. Meanwhile we also point out to the efficient enhancement over time of the atomic D abundance and the atomic D/H ratio (Fig.~\ref{grafH}) in the mantle. Hydrogen atoms are much more sticky and with much lower diffusion coefficient than hydrogen molecules (Strauss et al. \cite{10}). However, the growing abundance of atomic deuterium in the mantle even after $3\times10^8$ years is still too low to affect the isotoptic composition of molecules.

\end{document}